\documentclass[12pt,leqno]{article}

\usepackage{amsmath,amssymb,amsfonts}
\usepackage{graphicx}
\usepackage{booktabs}
\usepackage[numbers]{natbib}
\usepackage{hyperref}
\usepackage{geometry}
\usepackage{setspace}
\usepackage{caption}

\newcommand{\note}[1]{%
  \vspace{0.5\baselineskip}%
  \noindent\small\textit{Note:} #1%
}

\title{AI-Powered Energy Algorithmic Trading: Integrating Hidden Markov Models with Neural Networks}
\author{Tiago Monteiro\\
\small Northeastern University\\
\small Email: \texttt{monteiro.t@northeastern.edu}}

\begin{document}

\maketitle

\begin{abstract}
\noindent In quantitative finance, machine-learning methods are key for alpha generation. This study introduces a refreshing approach that combines Hidden Markov Models (HMM) and neural networks integrated with Black-Litterman portfolio optimization. The approach was tested during the COVID period (2019--2022), where it achieved an 83\% return with a Sharpe ratio of 0.77. Two risk models were incorporated to enhance risk management, particularly during the volatile periods. The methodology was implemented on the QuantConnect platform, which was chosen for its robust framework and experimental reproducibility. The system is designed to predict future price movements and includes a three-year warm-up period to ensure the proper use of the algorithm. It focuses on highly liquid, large-cap energy stocks to ensure stable and reliable results, while also accounting for broker payments. The dual-model alpha system utilizes log returns to select the optimal state based on historical performance. It combines state predictions with neural network outputs derived from historical data to generate trading signals. This study presents an in-depth examination of the trading system's architecture, data pre-processing, training, and performance. The full code and backtesting data are available under QuantConnect's terms.

\vspace{0.3\baselineskip}

\vspace{0.2\baselineskip}
\end{abstract}

\section{Introduction}

Algorithmic trading influences financial markets by enabling the rapid and exact execution of trading strategies beyond human capability \citep{dananjayan2023}. Machine learning integration has further reinvented this field by providing novel techniques for pattern recognition and predictive modeling \citep{afua2024}.

QuantConnect, chosen for this study, offers data access, back-testing, and a powerful algorithmic trading framework. This simplifies the development, testing, and deployment of trading strategies. The environment also ensures result replication, which is crucial for assessing the advanced trading algorithms.

This study outlines a novel methodology that fuses the HMM and neural networks to create a dual-model alpha generation system. The aim is to leverage the HMM's ability to capture temporal dependencies and market regimes, along with the power of neural networks to learn subtle patterns from historical price data \citep{oelschlager2020,elmorr2022}. The strategy uses Black-Litterman portfolio optimization combined with two risk management models.

The system employs a low-frequency buying strategy that is activated by converging signals from various AI models. A buy decision is made only when different model signals align.

The main objective is to develop an adaptable trading strategy that predicts price fluctuations and optimizes trading decisions. The results indicate the potential of this unified approach, with the algorithm achieving a 83\% return and a Sharpe ratio of 0.77 during the COVID period (2019--2022).

\section{Background}

HMMs, deep learning, and multi-model AI are crucial for optimizing algorithmic trading strategies. HMMs simulate systems with partially observable states, whereas deep learning utilizes neural networks to capture complex patterns. Multi-model AI integrates various machine learning models to improve robustness and accuracy \citep{giudici2024}.

The use of HMMs in financial markets has evolved from simple regime-switching models to advanced volatility modeling methods \citep{gorynin2017}. Deep learning has progressed from basic neural networks to sophisticated architectures, such as LSTM, CNN, and transformer-based models \citep{shiri2023}. Multi-model AI, employing ensemble methods, surpasses single ML algorithms in terms of robustness \citep{aelgani2023}. These theories form the basis for advanced trading strategies. HMMs provide insights into market regimes and volatility, deep learning models forecast price movements, and multi-model AI enhances the overall performance.

\subsection{Previous Research}

Previous research has highlighted the utility of HMMs in identifying market regimes and in volatility modeling. In addition, deep learning techniques have shown potential for time-series forecasting and sentiment analysis \citep{larabenitez2021}. Multi-model AI approaches, such as stacking and boosting, have improved prediction accuracy across various fields \citep{odegua2019}. HMM studies often use regime-switching models and probabilistic analysis, whereas deep learning focuses on neural network architectures and training methods. Multi-model AI research combines different algorithms through ensemble techniques to leverage their unique strengths.

\subsection{Challenges}

Despite their advantages, HMMs face challenges in capturing the market complexity. Simple HMMs may not ideally interpret market dynamics because of their inability to capture both immediate and extended trends \citep{oelschlager2021}. Deep learning models capture nonlinear relationships but require substantial data and computational resources. Although they often outperform traditional methods, their benefits can be modest and context-dependent, especially when data are limited or computational costs are high \citep{jiang2020}. Multi-model AI approaches offer enhanced robustness and predictive performance but are more complex to implement \citep{wang2021}. Variability in data, methods, and implementation can lead to inconsistencies, complicating the reproduction of results \citep{chen2022}.

\subsection{Research Gaps and Future Directions}

Researchers often downplay the benefits of HMMs, deep learning, and multi-model AI in trading strategies owing to the lack of standardized frameworks for strategy standardization and data pre-processing. This absence challenges the creation of single and multi-model systems \citep{afua2024b}. Bridging these gaps is essential to craft trading strategies capable of traversing complex financial markets. Future studies should explore integrating techniques to capture both linear and nonlinear market behavior, and investigate the impact of real-time market conditions on model performance \citep{bharath2023}.

\subsection{Tools and Hypotheses}

QuantConnect facilitates advanced trading by supporting multi-model deployment using the LEAN open-source engine. It standardizes data and algorithm development, improving efficiency and reproducibility, while addressing current limitations \citep{rashid2010}.

The execution of the hypotheses depends on incorporating HMM, deep learning, and multi-model AI to optimize algorithmic trading strategies. A comprehensive literature review clarifies each approach's advantages and limitations, forming the following research questions: How does integrating HMM, deep learning, and multi-model AI affect the effectiveness of algorithmic trading strategies? What computational challenges does this integration present, and how can they be mitigated? How does the unified model perform under market conditions compared with the S\&P 500, a standard benchmark for evaluating equity trading algorithms?

These hypotheses and research questions aim to develop an adaptive trading strategy that addresses the stated research gaps and takes advantage of HMM, deep learning, and multi-model AI strengths.

\section{Methodology}

\begin{figure}[t]
\centering
\includegraphics[width=0.85\textwidth]{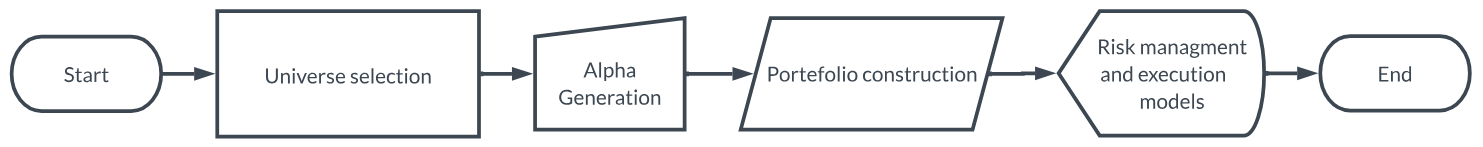}
\caption{Workflow of Algorithmic Trading System}
\label{fig:workflow}
\note{This figure illustrates the complete workflow from data collection through signal generation to portfolio execution.}
\end{figure}

This section explains how the dual-model system for generating trading signals was created using the HMM and neural networks. The QuantConnect platform was used in this project because it provides significant market data and streamlines strategy testing and deployment. Since 2013, it has been refined through over two million algorithms, ensuring reproducibility and transparency \citep{quantconnect2024}.

However, QuantConnect has limitations, such as data availability, which might lead to missing market scenarios required for training deep learning models. Future studies can use distributed computing to address these limitations. This study covers aspects such as model design, data management, training, signal generation, and portfolio strategies, as illustrated in figure \ref{fig:workflow}. A transparent and reproducible approach was used to confirm the reliability and success of the system.

Full code and backtesting data are available and licensed under the QuantConnect terms.

\subsection{Universe Selection}

This study used a dynamic universe with two filters to select tradable securities. The first filter evaluates liquidity through the dollar volume by selecting stocks with the highest total traded value. The second filter focuses on energy sector stocks based on market capitalization and identifies the top 20 companies in this sector. Larger companies typically exhibit more stable and predictable performance, beneficial for analysis \citep{sauberschwarz2017}. This way, this dynamic universe prioritizes liquidity and market capitalization. Establishing a solid foundation for scientific models to operate effectively without artificial complexity \citep{rakovic2018}.

\subsection{Alpha Generation}

\subsubsection{Model Architecture}

The system leverages the fusion of an HMM and a neural network based on PyTorch to examine the historical price data, generate trading insights, and facilitate their integration.

\paragraph{Hidden Markov Model}

\begin{figure}[t]
\centering
\includegraphics[width=0.85\textwidth]{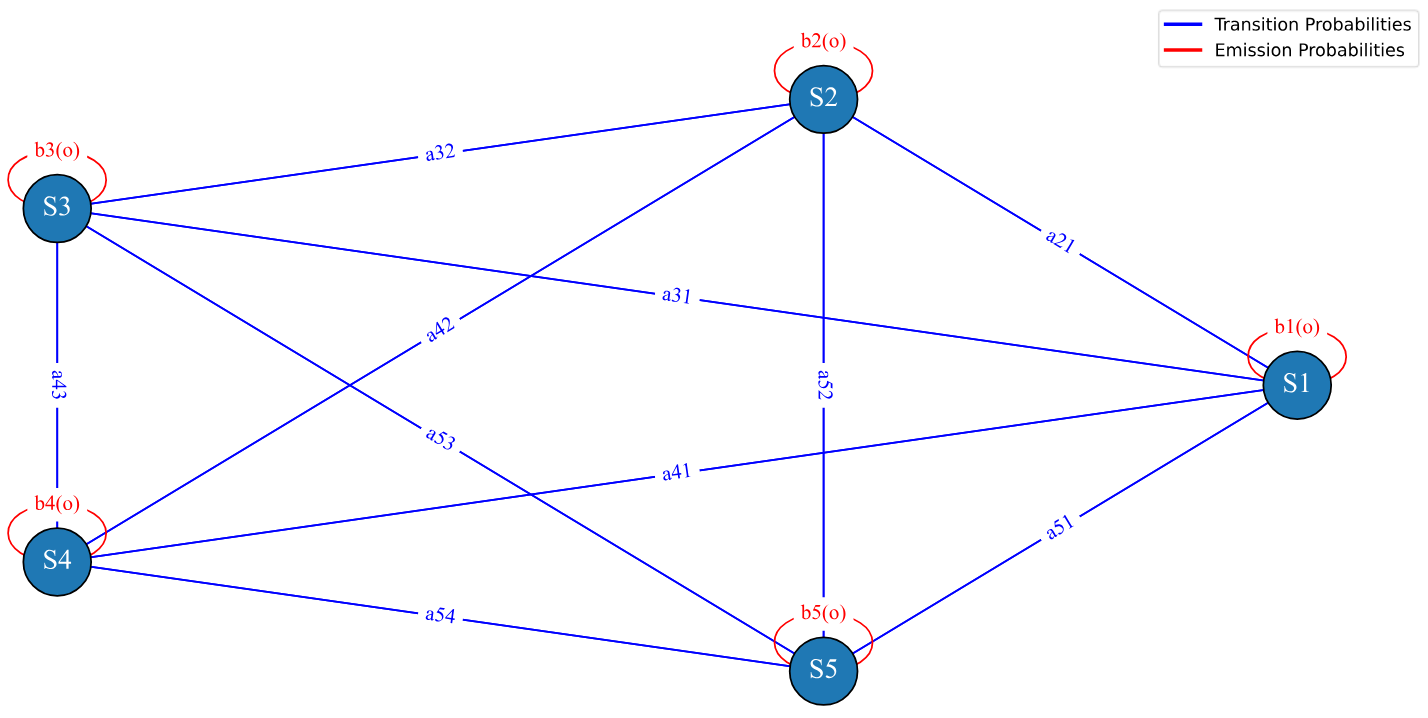}
\caption{HMM Architecture for Simulating Stock Price Fluctuations}
\label{fig:hmm}
\note{The model uses five hidden states to capture different market regimes including bull, bear, and transitional phases.}
\end{figure}

The HMM is a stochastic framework for simulating stock price fluctuations (figure \ref{fig:hmm}). An HMM is formally defined by the tuple $\lambda = (S, O, A, B, \pi)$ where:
\begin{itemize}
\item $S = \{s_1, s_2, \ldots, s_N\}$ is the set of $N$ hidden states (in this study, $N=5$)
\item $O = \{o_1, o_2, \ldots, o_T\}$ is the sequence of observations (log returns)
\item $A = \{a_{ij}\}$ is the state transition probability matrix, where
\begin{equation}
a_{ij} = P(s_t = j | s_{t-1} = i), \quad \sum_{j=1}^{N} a_{ij} = 1
\end{equation}
\item $B = \{b_j(o_t)\}$ is the observation probability distribution
\item $\pi = \{\pi_i\}$ is the initial state distribution, where $\pi_i = P(s_1 = i)$
\end{itemize}

The model is trained on log returns, calculated as:
\begin{equation}
r_t = \ln\left(\frac{P_t}{P_{t-1}}\right)
\end{equation}
where $P_t$ is the closing price at time $t$.

For a Gaussian HMM with full covariance, the emission probability is:
\begin{equation}
b_j(o_t) = \mathcal{N}(o_t; \mu_j, \Sigma_j) = \frac{1}{\sqrt{(2\pi)^d |\Sigma_j|}} \exp\left(-\frac{1}{2}(o_t - \mu_j)^T \Sigma_j^{-1} (o_t - \mu_j)\right)
\end{equation}
where $\mu_j$ and $\Sigma_j$ are the mean vector and covariance matrix for state $j$.

The optimal state sequence is determined using the Viterbi algorithm, which finds:
\begin{equation}
s^* = \arg\max_{s_{1:T}} P(s_{1:T} | o_{1:T}, \lambda)
\end{equation}

The best state for prediction is selected based on historical mean return:
\begin{equation}
s_{\text{best}} = \arg\max_j \mathbb{E}[r_t | s_t = j]
\end{equation}

The model runs for up to 10 iterations during fitting to ensure convergence while preserving computational efficiency. The five hidden states represent various market conditions, such as bull, bear, and transitional phases, offering a broader market view.

The HMM can overfit past data, affecting the prediction accuracy with respect to abrupt market changes. Future research should explore regularization techniques and hybrid models that combine HMM with unsupervised machine learning to navigate market transitions.

\paragraph{PyTorch-Based Neural Network}

\begin{figure}[t]
\centering
\includegraphics[width=0.8\textwidth]{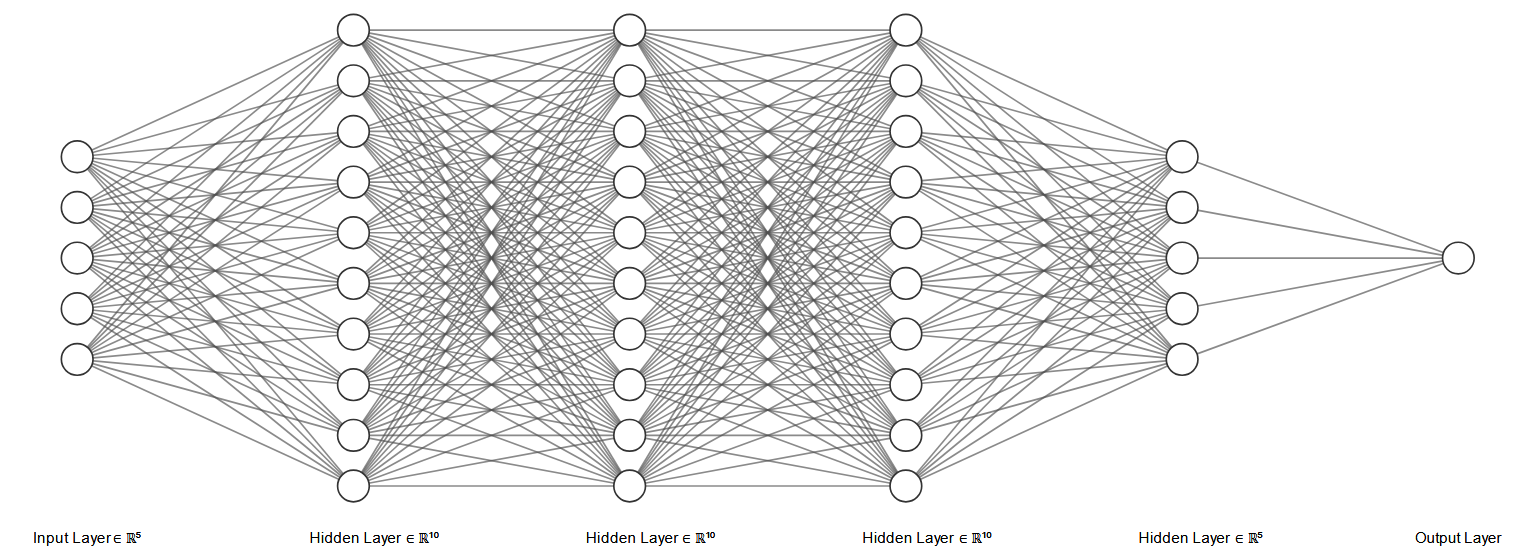}
\caption{Neural Network Architecture for Detecting Complex Patterns in Historical Price Data}
\label{fig:nn}
\note{The feedforward network consists of four hidden layers with ReLU activations, trained using the Adam optimizer.}
\end{figure}

The neural network aims to detect complex patterns in the historical price data, as shown in figure \ref{fig:nn}. The architecture is a fully connected feedforward neural network with the following mathematical formulation:

The input layer receives a feature vector $\mathbf{x} \in \mathbb{R}^5$ representing five consecutive historical price differences:
\begin{equation}
\mathbf{x} = [P_{t-4} - P_{t-5}, P_{t-3} - P_{t-4}, \ldots, P_t - P_{t-1}]^T
\end{equation}

The network architecture consists of four hidden layers with ReLU activation functions and one output layer:
\begin{align}
\mathbf{h}_1 &= \text{ReLU}(W_1 \mathbf{x} + \mathbf{b}_1), \quad \mathbf{h}_1 \in \mathbb{R}^{10} \\
\mathbf{h}_2 &= \text{ReLU}(W_2 \mathbf{h}_1 + \mathbf{b}_2), \quad \mathbf{h}_2 \in \mathbb{R}^{10} \\
\mathbf{h}_3 &= \text{ReLU}(W_3 \mathbf{h}_2 + \mathbf{b}_3), \quad \mathbf{h}_3 \in \mathbb{R}^{10} \\
\mathbf{h}_4 &= \text{ReLU}(W_4 \mathbf{h}_3 + \mathbf{b}_4), \quad \mathbf{h}_4 \in \mathbb{R}^{5} \\
\hat{y} &= W_5 \mathbf{h}_4 + b_5, \quad \hat{y} \in \mathbb{R}
\end{align}
where $W_i$ and $\mathbf{b}_i$ are the weight matrices and bias vectors for layer $i$, and $\text{ReLU}(z) = \max(0, z)$ is the Rectified Linear Unit activation function.

The network is trained to minimize the mean squared error (MSE) loss function:
\begin{equation}
\mathcal{L}(\theta) = \frac{1}{n} \sum_{i=1}^{n} (y_i - \hat{y}_i)^2
\end{equation}
where $\theta = \{W_1, \mathbf{b}_1, \ldots, W_5, b_5\}$ represents all trainable parameters, $y_i$ is the actual price change, and $\hat{y}_i$ is the predicted price change.

The Adam optimizer is employed for parameter updates with adaptive learning rates:
\begin{align}
m_t &= \beta_1 m_{t-1} + (1 - \beta_1) \nabla_\theta \mathcal{L}_t \\
v_t &= \beta_2 v_{t-1} + (1 - \beta_2) (\nabla_\theta \mathcal{L}_t)^2 \\
\hat{m}_t &= \frac{m_t}{1 - \beta_1^t}, \quad \hat{v}_t = \frac{v_t}{1 - \beta_2^t} \\
\theta_{t+1} &= \theta_t - \alpha \frac{\hat{m}_t}{\sqrt{\hat{v}_t} + \epsilon}
\end{align}
where $\alpha = 0.001$ is the learning rate, $\beta_1 = 0.9$ and $\beta_2 = 0.999$ are the exponential decay rates, and $\epsilon = 10^{-8}$ is a small constant for numerical stability. The model was trained for five epochs to balance training time and risk of overfitting.

The complete network architecture can be represented as:
\begin{equation}
f_{NN}: \mathbb{R}^5 \xrightarrow{W_1} \mathbb{R}^{10} \xrightarrow{W_2} \mathbb{R}^{10} \xrightarrow{W_3} \mathbb{R}^{10} \xrightarrow{W_4} \mathbb{R}^{5} \xrightarrow{W_5} \mathbb{R}
\end{equation}

Neural networks can encounter challenges in hyperparameter tuning and overfitting. Overcoming these challenges can be achieved by cross-validation and automated hyperparameter optimization.

\subsubsection{Data Management}

Effective data management is crucial for optimizing the model's forecasting performance. The algorithm uses a rolling window approach to update historical price data continuously. At each time step $t$, a window of size $k$ is maintained:
\begin{equation}
\mathcal{W}_t = \{P_{t-k+1}, P_{t-k+2}, \ldots, P_t\}
\end{equation}

Feature extraction differs between the two models to leverage their respective strengths:

For the HMM, log returns are computed as:
\begin{equation}
r_t = \ln\left(\frac{P_t}{P_{t-1}}\right)
\end{equation}

For the neural network, price differences are used as features:
\begin{equation}
\Delta P_t = P_t - P_{t-1}
\end{equation}

Data normalization is applied to ensure numerical stability:
\begin{equation}
x_{\text{norm}} = \frac{x - \mu}{\sigma}
\end{equation}
where $\mu$ and $\sigma$ are the mean and standard deviation computed over the rolling window.

Data integrity was ensured through validation, adding only trade bars with valid closing prices, and using QuantConnect's infrastructure for data consistency. The validation criterion requires:
\begin{equation}
P_t > 0 \quad \text{and} \quad |r_t| < \tau
\end{equation}
where $\tau$ is a threshold for detecting anomalous returns (e.g., $\tau = 0.5$ representing 50\% change).

Regular retraining occurs at frequency $f$ (measured in days) to ensure the models adapt to evolving market conditions:
\begin{equation}
\text{Retrain at } t \text{ if } t \bmod f = 0
\end{equation}

This approach ensures that models have fresh data for predictions while maintaining computational efficiency via error handling and robust data validation across market conditions.

\subsubsection{Models Training}

HMM training uses a Gaussian HMM with several components to capture and adapt to different market regimes. Neural network training occurs in mini-batches over multiple epochs and involves forward propagation, loss calculation, backpropagation, and parameter updates. This repeatable process enables the network to learn from data, refine parameters, and improve accuracy. A three-year preparation period provides historical data to help models recognize patterns before making predictions.

Cross-validation and regularization techniques are essential for preventing underfitting and overfitting. Incorporating validation metrics and evaluation methods further enhances model performance assessment.

\subsubsection{Insight Generation}

The dual-model system generates trading insights by integrating predictions from both the HMM and neural network through a consensus-based decision framework.

The HMM produces a state prediction for the next time step using the forward algorithm:
\begin{equation}
\hat{s}_{t+1} = \arg\max_{j \in S} P(s_{t+1} = j | o_{1:t}, \lambda)
\end{equation}

For each state $j$, the expected return is computed from historical data:
\begin{equation}
\mathbb{E}[r | s = j] = \frac{1}{|T_j|} \sum_{t \in T_j} r_t
\end{equation}
where $T_j = \{t : s_t = j\}$ is the set of time steps when the system was in state $j$.

The Neural Network predicts the next price change:
\begin{equation}
\hat{P}_{t+1} = P_t + f_{NN}(\mathbf{x}_t; \theta)
\end{equation}
where $f_{NN}$ is the trained neural network function with parameters $\theta$.

The trading signal is generated by combining both predictions through the following decision rule:
\begin{equation}
\text{Signal}_t = 
\begin{cases}
+1 & \text{(Buy) if } \mathbb{E}[r | s = \hat{s}_{t+1}] > \epsilon \text{ AND } \hat{P}_{t+1} > P_t \\
-1 & \text{(Sell) if } \mathbb{E}[r | s = \hat{s}_{t+1}] < -\epsilon \text{ AND } \hat{P}_{t+1} < P_t \\
0 & \text{(Hold) otherwise}
\end{cases}
\end{equation}
where $\epsilon$ is a threshold parameter that filters out weak signals (typically $\epsilon = 0.001$ or 0.1\%).

The confidence level of the signal can be quantified as:
\begin{equation}
\text{Confidence}_t = \left|\mathbb{E}[r | s = \hat{s}_{t+1}]\right| \times P(s_{t+1} = \hat{s}_{t+1} | o_{1:t})
\end{equation}

This consensus mechanism ensures that a trading signal is only generated when both models agree on the direction of price movement, thereby reducing false signals and improving overall strategy reliability. The approach can be formalized as:
\begin{equation}
\text{Execute}_t = 
\begin{cases}
\text{True} & \text{if } \text{sign}(\mathbb{E}[r | s = \hat{s}_{t+1}]) = \text{sign}(\hat{P}_{t+1} - P_t) \\
\text{False} & \text{otherwise}
\end{cases}
\end{equation}

Possible conflicts between model predictions can be addressed using ensemble techniques, such as stacking, weighted voting, and gradient boosting, which enhance prediction accuracy and resilience by leveraging the strengths of each model.

\subsection{Portfolio Construction}

Proper portfolio construction is crucial for the success of the algorithm. The Black-Litterman model was chosen for its innovative approach, combining expected returns and risk. Created by Fischer Black and Robert Litterman, this model merges investor views with market equilibrium for better asset allocation \citep{cayirli2019}.

The Black-Litterman model combines market equilibrium returns with investor views to produce posterior expected returns. The model starts with the market equilibrium return vector $\Pi$, derived from market capitalization weights:
\begin{equation}
\Pi = \lambda \Sigma w_{\text{mkt}}
\end{equation}
where $\lambda$ is the risk aversion coefficient, $\Sigma$ is the covariance matrix of asset returns, and $w_{\text{mkt}}$ is the market capitalization weight vector.

The investor views are expressed through a picking matrix $P$ and view portfolio returns $Q$, with uncertainty in the views captured by the diagonal matrix $\Omega$. The posterior expected return is then computed as:
\begin{equation}
\mathbb{E}[R] = \left[(\tau\Sigma)^{-1} + P^T\Omega^{-1}P\right]^{-1}\left[(\tau\Sigma)^{-1}\Pi + P^T\Omega^{-1}Q\right]
\end{equation}
where $\tau$ is a scalar parameter representing the uncertainty in the prior (typically $\tau \in [0.01, 0.05]$).

The posterior covariance matrix is:
\begin{equation}
\Sigma_{\text{post}} = \Sigma + \left[(\tau\Sigma)^{-1} + P^T\Omega^{-1}P\right]^{-1}
\end{equation}

Optimal portfolio weights are determined by maximizing the expected utility:
\begin{equation}
w^* = \arg\max_w \left\{w^T\mathbb{E}[R] - \frac{\lambda}{2}w^T\Sigma_{\text{post}}w\right\}
\end{equation}

This yields the optimal weight vector:
\begin{equation}
w^* = \frac{1}{\lambda}(\Sigma_{\text{post}})^{-1}\mathbb{E}[R]
\end{equation}
subject to the constraint $\sum_{i=1}^N w_i = 1$ (fully invested portfolio).

The expected portfolio return and variance are:
\begin{align}
R_p &= \sum_{i=1}^{N} w_i^* R_i = (w^*)^T \mathbb{E}[R] \\
\sigma_p^2 &= (w^*)^T \Sigma_{\text{post}} w^*
\end{align}

This model increases algorithm modularity, separating alpha generation from portfolio management \citep{tzang2020}. Unlike signal-based portfolio construction, it reduces the probability of inaccurate signals, preventing inefficient management in achieving the optimal Sharpe ratio. The Black-Litterman model improves system reliability by clearly dividing alpha generation and portfolio management. It directs alpha model insights into risk management and execution models and minimizes unnecessary trades and risks.

\subsection{Risk Management and Execution Models}

This study integrates two risk management models to enhance portfolio robustness. The selected models are the Maximum Drawdown Percent per security and the trailing-stop risk management model.

\paragraph{Maximum Drawdown Model}

The maximum drawdown (MDD) for a security measures the largest peak-to-trough decline in portfolio value. For a given security $i$, the drawdown at time $t$ is:
\begin{equation}
DD_i(t) = \frac{\max_{s \in [0,t]} V_i(s) - V_i(t)}{\max_{s \in [0,t]} V_i(s)}
\end{equation}
where $V_i(t)$ is the value of security $i$ at time $t$.

The maximum drawdown over the entire period is:
\begin{equation}
\text{MDD}_i = \max_{t \in [0,T]} DD_i(t)
\end{equation}

A position in security $i$ is automatically closed if:
\begin{equation}
DD_i(t) > \text{MDD}_{\text{threshold}}
\end{equation}
where $\text{MDD}_{\text{threshold}}$ is typically set between 10\% and 20\%, depending on risk tolerance. This model protects investor capital by automatically exiting positions with excessive losses, thereby ensuring disciplined risk controls.

\paragraph{Trailing Stop Model}

The trailing-stop model dynamically adjusts stop-loss levels to capture gains while protecting against losses. For a long position in security $i$, the trailing stop price is updated as:
\begin{equation}
\text{Stop}_i(t) = \max\left(P_i(t) \times (1 - \delta), \text{Stop}_i(t-1)\right)
\end{equation}
where $\delta$ is the trailing stop percentage (typically $\delta \in [0.05, 0.15]$), and $P_i(t)$ is the current price.

The position is closed if:
\begin{equation}
P_i(t) \leq \text{Stop}_i(t)
\end{equation}

For short positions, the trailing stop is computed as:
\begin{equation}
\text{Stop}_i(t) = \min\left(P_i(t) \times (1 + \delta), \text{Stop}_i(t-1)\right)
\end{equation}

The trailing-stop model locks in profits in trending markets while providing downside protection. The expected profit captured by the trailing stop can be approximated as:
\begin{equation}
\mathbb{E}[\text{Profit}] \approx \max_{t \in [t_{\text{entry}}, t_{\text{exit}}]} \left[P_i(t) - P_i(t_{\text{entry}})\right] \times (1 - \delta)
\end{equation}

These risk management models, combined with the lean engine's execution model, ensure timely trade execution and capitalize on market opportunities. This approach supports robust prediction models and effective asset allocation, balancing the alpha generation with disciplined risk management. The combined risk-adjusted return can be expressed as:
\begin{equation}
R_{\text{adj}} = R_p - \lambda_{\text{risk}} \times \text{Risk}_{\text{penalty}}
\end{equation}
where $\lambda_{\text{risk}}$ is the risk aversion parameter and $\text{Risk}_{\text{penalty}}$ captures drawdown and volatility concerns.

\section{Results and Discussion}

\subsection{Key Statistics}

\begin{table}[t]
\centering
\caption{Key Trading Metrics}
\begin{tabular*}{\textwidth}{@{\extracolsep\fill}lc}
\toprule
\textbf{Metric} & \textbf{Value} \\
\midrule
\multicolumn{2}{l}{\textit{Performance Metrics}} \\
\quad Runtime Days & 1,096 \\
\quad Compounded Annual Growth Rate (CAGR) & 22.2\% \\
\quad Sharpe Ratio (Probabilistic) & 30\% \\
\quad Sortino Ratio & 0.6 \\
\quad Information Ratio & $-0.1$ \\
\addlinespace
\multicolumn{2}{l}{\textit{Risk Metrics}} \\
\quad Maximum Drawdown & 17.1\% \\
\quad Portfolio Turnover & 3.29\% \\
\addlinespace
\multicolumn{2}{l}{\textit{Capacity}} \\
\quad Optimal Investment Level (USD) & \$10M \\
\bottomrule
\end{tabular*}
\label{tab:key_metrics}
\note{Performance metrics cover the full backtesting period from January 2019 to January 2022, encompassing the COVID-19 market volatility.}
\end{table}

This table presents the essential metrics used to evaluate the performance and risk of an investment portfolio. It covers Runtime Days over a period of 1096 days to assess performance sustainability and volatility. The Drawdown indicates a maximum loss from peak to trough of 17.1\%, reflecting the risk of potential declines. A Portfolio Turnover of 3.29\% signifies trading frequency, with lower figures suggesting a more passive management strategy aimed at minimizing transaction costs.

Risk-adjusted performance was assessed using several ratios, as summarized in table \ref{tab:key_metrics}. These metrics provide a comprehensive evaluation of the trading strategy's effectiveness:

The Sharpe Ratio measures risk-adjusted returns:
\begin{equation}
S = \frac{\mathbb{E}[R_p - R_f]}{\sigma_p}
\end{equation}
where $R_p$ is the portfolio return, $R_f$ is the risk-free rate, and $\sigma_p$ is the portfolio standard deviation. The probabilistic Sharpe ratio at 41\% quantifies returns relative to risk, with higher ratios denoting more efficient risk management.

The Compounded Annual Growth Rate (CAGR) at 22.2\% represents the geometric mean return:
\begin{equation}
\text{CAGR} = \left(\frac{V_{\text{final}}}{V_{\text{initial}}}\right)^{\frac{1}{n}} - 1
\end{equation}
where $V_{\text{final}} = \$182,761$, $V_{\text{initial}} = \$100,000$, and $n = 1096/365 = 3$ years.

The Sortino Ratio of 0.6 focuses on downside risk:
\begin{equation}
S_{\text{sortino}} = \frac{\mathbb{E}[R_p - R_f]}{\sigma_{\text{downside}}}
\end{equation}
where $\sigma_{\text{downside}} = \sqrt{\mathbb{E}[\min(R_p - R_f, 0)^2]}$ measures downside deviation.

The Information Ratio of $-0.1$ compares performance to a benchmark:
\begin{equation}
IR = \frac{\mathbb{E}[R_p - R_b]}{\sigma_{R_p - R_b}}
\end{equation}
where $R_b$ is the benchmark return (S\&P 500) and $\sigma_{R_p - R_b}$ is the tracking error.

Further details include the Capacity (USD) at \$10 million, which represents the optimal investment level for strategy sustainability and achieving target returns.

\subsection{Detailed Trading Metrics}

\begin{table}[t]
\centering
\caption{Detailed Trading Metrics and Performance Indicators}
\begin{tabular*}{\textwidth}{@{\extracolsep\fill}lc}
\toprule
\textbf{Metric} & \textbf{Value} \\
\midrule
\multicolumn{2}{l}{\textit{Trading Activity}} \\
\quad Total Orders Executed & 40 \\
\quad Win Rate & 60\% \\
\quad Loss Rate & 40\% \\
\quad Average Win & 7.70\% \\
\quad Average Loss & $-3.29\%$ \\
\quad Profit-Loss Ratio & 2.34 \\
\addlinespace
\multicolumn{2}{l}{\textit{Portfolio Performance}} \\
\quad Initial Equity & \$100,000 \\
\quad Final Equity & \$182,761 \\
\quad Total Return & 82.76\% \\
\quad Sharpe Ratio & 0.77 \\
\quad Treynor Ratio & 0.924 \\
\addlinespace
\multicolumn{2}{l}{\textit{Risk-Adjusted Returns}} \\
\quad Alpha & 0.13 \\
\quad Beta & 0.175 \\
\quad Annual Standard Deviation & 0.21 \\
\quad Annual Variance & 0.044 \\
\quad Tracking Error & 0.256 \\
\addlinespace
\multicolumn{2}{l}{\textit{Transaction Costs}} \\
\quad Total Broker Fees & \$972 \\
\bottomrule
\end{tabular*}
\label{tab:detailed_metrics}
\note{The low trading frequency (40 total orders) reflects the conservative, signal-based approach of the dual-model system.}
\end{table}

The metrics evaluate trading outcomes over a specified period, revealing a relatively low frequency, with only 40 trades. An average win rate of 7.70\% and an average loss of $-3.29\%$, with a 60\% success rate, highlight the effectiveness of the strategy. The initial equity of \$100,000 increased to \$182,761.12, indicating substantial gain. A Sharpe Ratio of 0.77 demonstrates a risk-return trade-off.

An alpha of 0.13 suggests modest active returns, and a beta of 0.175 indicates lower market volatility than broader indices, as detailed in table \ref{tab:detailed_metrics}. An annual standard deviation of 0.21 and a variance of 0.044 quantified the return variability. A tracking error of 0.256 indicates how the portfolio's returns deviate from the benchmark, whereas a Treynor Ratio of 0.951 indicates good risk-adjusted returns. Fees amounting to \$971.62, are relatively small compared to growth in equity.

These data indicate a robust trading strategy that can effectively manage risk. Optimizing hyperparameters and preprocessing data can enhance model performance and improve results. However, extensive Monte Carlo simulations are necessary to ensure model reliability. This reinforces the importance of rigorous data analysis and strategy refinement in trading.

\subsection{Cumulative Returns}

\begin{figure}[t]
\centering
\includegraphics[width=\textwidth]{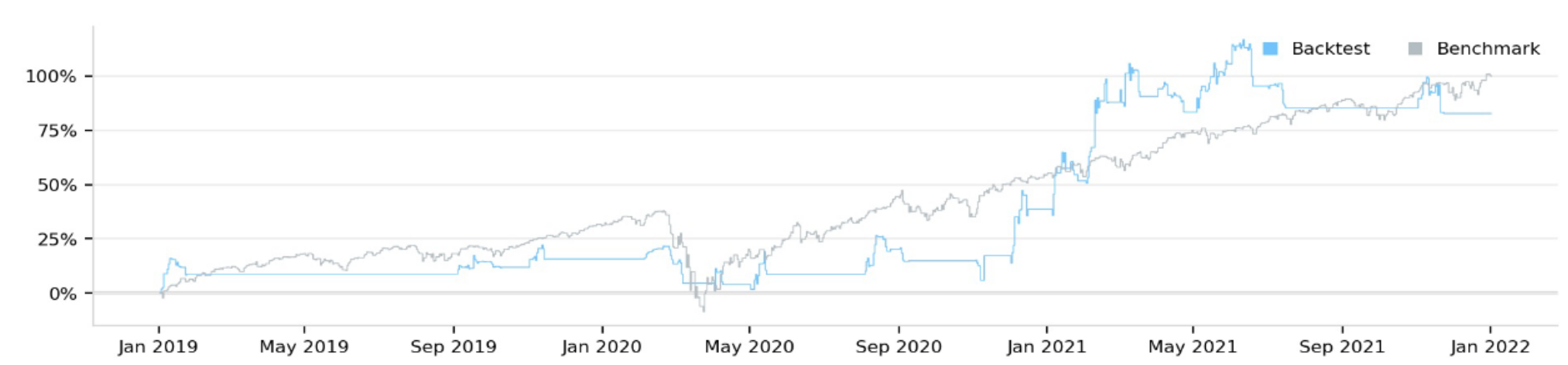}
\caption{Cumulative Returns Comparison (January 2019--January 2022)}
\label{fig:cumulative}
\note{The light blue line represents the backtested strategy while the gray line shows the S\&P 500 benchmark. Notable outperformance occurred during the COVID-19 market volatility in May 2020 and September 2021.}
\end{figure}

The Cumulative Returns chart illustrates the performance comparison of a backtested investment strategy against a benchmark from January 2019 to January 2022. The x-axis represents the timeline, marked in months, while the y-axis quantifies cumulative returns in percentage terms. The backtested strategy is depicted by a light blue line, whereas the benchmark (presumed to be the S\&P 500) is shown in gray.

Throughout the observation period, the strategy generally aligns closely with the benchmark, indicating a strong market correlation. Notably, the strategy outperformed the benchmark at specific points, such as during the onset of the COVID-19 pandemic in May 2020 and September 2021, as illustrated in figure \ref{fig:cumulative}. These surges suggest effective strategy adjustments in response to market volatility.

Despite some instances of notable growth, the strategy often mirrors the benchmark's movements, highlighting its dependence on broader market trends rather than consistently outperforming it. During the COVID-19 crisis, the strategy demonstrated resilience, maintaining its value better than the benchmark and taking advantage of the recovery phase, which is indicative of adept risk management and opportunity utilization.

\subsection{Advantages of the Approach}

Utilizing the HMM alongside neural networks offers notable advantages for predictive modeling in financial markets. HMMs capture temporal dependencies crucial for sequential data like stock prices, while neural networks excel in recognizing complex patterns and relationships \citep{alwateer2023,qamar2023}. By combining these approaches, a more complete and accurate prediction model is achieved, leveraging the strengths of both strategies. This hybrid approach enhances the robustness by reducing the biases inherent in any single model, resulting in more reliable trading signals. Risk management strategies also play a crucial role in guarding against potential losses and improving the overall portfolio stability. The adaptability gained from merging HMMs and neural networks is essential for maintaining a high performance across different market conditions.

\subsection{Future Research Directions}

Future research should explore various approaches to enhance dual-model methodology by combining HMMs and neural networks.

\paragraph{Wavelet Transform for Enhanced Market Pattern Recognition}

One potential approach is to apply a wavelet transform to the pre-processing strategy of the data. Wavelet transforms capture both time and frequency domain information, improving the potential of the model to recognize patterns in price data, which is crucial to ensuring its success \citep{kwon2022}. This is vital as it accounts for the dynamic nature of the market, making it more effective than static methods such as Pearson's correlation coefficient.

\paragraph{Monte Carlo Simulations for Model Generalizability}

The use of extensive Monte Carlo simulations across diverse temporal frames is another promising area of research. These simulations help analyze the generalization capabilities of the model, evaluating its reliability and identifying possible cases of overfitting or incorrect market predictions \citep{zillich2009}. This ensures the stability and accuracy of the algorithmic trading system in a real trading scenario.

\paragraph{Hyperparameter Tuning and Ensemble Methods}

Employing novel neural network architectures along with automated hyperparameter fine-tuning can enhance forecasting capabilities by finding the optimal architecture for a specific dataset \citep{wu2024}. Additionally, incorporating an ensemble machine learning method for HMM and neural network systems can enable the model to adapt dynamically to market fluctuations, improving overall alpha generation performance \citep{ong2011}.

\paragraph{Diversifying Equities for Growth and Stability}

Increasing sector diversification is essential to enhance portfolio resilience. For example, combining the 25 largest stocks from high-potential sectors such as technology, biomedicine, pharmaceuticals, and energy can enhance diversification and access growth alternatives. This approach helps offset sector-specific risks, leverages the strengths of each sector, and theoretically leads to more stable returns.

\subsection{QuantConnect for Reproducibility}

QuantConnect was chosen for its powerful framework and user-friendly design, which simplified research replication. Its extensive access to data assures precise training and testing of the model. The integrated lean open-source algorithmic trading framework allows the implementation and testing of complex strategies, accelerating the development and validation of innovative models. This platform enhances research replicability, enabling other researchers and practitioners to validate and extend the outcomes. Which is essential for the advancement of scientific research.

\section{Conclusion}

The combined use of the HMM and neural networks in a dual-model approach has shown success in enhancing the accuracy and resilience of trading signals. During the COVID period (2019--2022), the dual-model system achieved an 83\% return and a Sharpe ratio of 0.77, highlighting its practical potential.

A three-year preparation period provides historical data to help models recognize patterns and understand market dynamics before making predictions. However, replicating these outcomes in complex financial markets is harder. These results suggest that combining the HMM and neural networks can strengthen algorithmic trading by generating more precise and reliable signals. Additionally, integration of multiple risk models improves the robustness of the system.

This study underscores the benefits of combining HMMs and neural networks for signal trading. Future research could explore wavelet transform pre-processing, extensive Monte Carlo simulations, and adaptive machine learning methods to further improve model performance. Using QuantConnect can enhance the reproducibility of results.

The full code and backtesting data are available and licensed under the QuantConnect terms.

\section*{Acknowledgments}

The author acknowledges the QuantConnect platform for providing the infrastructure and data access necessary for this research.

\section*{Conflict of interest}

The author declares no conflicts of interest in this paper.

\bibliographystyle{plainnat}
\bibliography{references}

@article{dananjayan2023,
  author = {Dananjayan, M. P. and Gopakumar, S. and Narayanasamy, P.},
  title = {Unleashing the algorithmic frontier: {Navigating} the impact of algo trading on investor portfolios},
  journal = {Journal of Information Technology Teaching Cases},
  year = {2023},
  doi = {10.1177/20438869231189519},
  url = {https://doi.org/10.1177/20438869231189519}
}

@article{afua2024,
  author = {Afua, W. and Ajayi-Nifise, N. A. O. and Bello, G. and others},
  title = {Algorithmic Trading and {AI}: {A} Review of Strategies and Market Impact},
  journal = {World Journal of Advanced Engineering Technology and Sciences},
  year = {2024},
  volume = {11},
  pages = {258--267},
  doi = {10.30574/wjaets.2024.11.1.0054},
  url = {https://doi.org/10.30574/wjaets.2024.11.1.0054}
}

@article{oelschlager2020,
  author = {Oelschl{\"a}ger, L. and Adam, T.},
  title = {Detecting bearish and bullish markets in financial time series using hierarchical hidden {Markov} models},
  journal = {arXiv},
  year = {2020},
  doi = {10.48550/arxiv.2007.14874},
  url = {https://doi.org/10.48550/arxiv.2007.14874}
}

@incollection{elmorr2022,
  author = {El Morr, C. and Jammal, M. and Ali-Hassan, H. and others},
  title = {Neural Networks},
  booktitle = {International Series in Management Science/Operations Research},
  year = {2022},
  pages = {319--360},
  doi = {10.1007/978-3-031-16990-8_11},
  url = {https://doi.org/10.1007/978-3-031-16990-8_11}
}

@article{giudici2024,
  author = {Giudici, P. and Raffinetti, E. and Riani, M.},
  title = {Robust machine learning models: linear and nonlinear},
  journal = {International Journal of Data Science and Analytics},
  year = {2024},
  doi = {10.1007/s41060-024-00512-1},
  url = {https://doi.org/10.1007/s41060-024-00512-1}
}

@article{gorynin2017,
  author = {Gorynin, I. and Monfrini, E. and Pieczynski, W.},
  title = {Pairwise {Markov} models for stock index forecasting},
  journal = {Zenodo},
  year = {2017},
  doi = {10.23919/eusipco.2017.8081568},
  url = {https://doi.org/10.23919/eusipco.2017.8081568}
}

@article{shiri2023,
  author = {Shiri, F. M. and Perumal, T. and Mustapha, N. and others},
  title = {A Comprehensive Overview and Comparative Analysis on Deep Learning Models: {CNN}, {RNN}, {LSTM}, {GRU}},
  journal = {arXiv},
  year = {2023},
  url = {https://arxiv.org/abs/2305.17473}
}

@article{aelgani2023,
  author = {Aelgani, V. and Vadlakonda, D.},
  title = {Explainable Artificial Intelligence based Ensemble Machine Learning for Ovarian Cancer Stratification using Electronic Health Records},
  journal = {International Journal of Recent Innovation Trends in Computing and Communication},
  year = {2023},
  volume = {11},
  pages = {78--84},
  doi = {10.17762/ijritcc.v11i7.7832},
  url = {https://doi.org/10.17762/ijritcc.v11i7.7832}
}

@article{larabenitez2021,
  author = {Lara-Benitez, P. and Carranza-Garcia, M. and Riquelme, J. C.},
  title = {An Experimental Review on Deep Learning Architectures for Time Series Forecasting},
  journal = {International Journal of Neural Systems},
  year = {2021},
  volume = {31},
  pages = {2130001},
  doi = {10.1142/s0129065721300011},
  url = {https://doi.org/10.1142/s0129065721300011}
}

@inproceedings{odegua2019,
  author = {Odegua, R.},
  title = {An Empirical Study of Ensemble Techniques (Bagging, Boosting and Stacking)},
  booktitle = {Proceedings of the Conference on Deep Learning, IndabaX},
  year = {2019},
  url = {https://www.researchgate.net/publication/338681876}
}

@article{oelschlager2021,
  author = {Oelschl{\"a}ger, L. and Adam, T.},
  title = {Detecting bearish and bullish markets in financial time series using hierarchical hidden {Markov} models},
  journal = {Statistical Modelling},
  year = {2021},
  volume = {1471082X2110340},
  doi = {10.1177/1471082x211034048},
  url = {https://doi.org/10.1177/1471082x211034048}
}

@article{jiang2020,
  author = {Jiang, W.},
  title = {Time series classification: nearest neighbor versus deep learning models},
  journal = {SN Applied Sciences},
  year = {2020},
  volume = {2},
  doi = {10.1007/s42452-020-2506-9},
  url = {https://doi.org/10.1007/s42452-020-2506-9}
}

@article{wang2021,
  author = {Wang, J. and Li, Q. and Zeng, B.},
  title = {Multi-layer cooperative combined forecasting system for short-term wind speed forecasting},
  journal = {Sustainable Energy Technologies and Assessments},
  year = {2021},
  volume = {43},
  pages = {100946},
  doi = {10.1016/j.seta.2020.100946},
  url = {https://doi.org/10.1016/j.seta.2020.100946}
}

@article{chen2022,
  author = {Chen, W.-B. and Li, X.-Y. and Kang, R.},
  title = {Integration for degradation analysis with multi-source {ADT} datasets considering dataset discrepancies and epistemic uncertainties},
  journal = {Reliability Engineering \& System Safety},
  year = {2022},
  volume = {222},
  pages = {108430},
  doi = {10.1016/j.ress.2022.108430},
  url = {https://doi.org/10.1016/j.ress.2022.108430}
}

@article{afua2024b,
  author = {Afua, W. and Ajayi-Nifise, N. A. O. and Bello, G. and others},
  title = {Algorithmic Trading and {AI}: {A} Review of Strategies and Market Impact},
  journal = {World Journal of Advanced Engineering Technology and Sciences},
  year = {2024},
  volume = {11},
  pages = {258--267},
  doi = {10.30574/wjaets.2024.11.1.0054},
  url = {https://doi.org/10.30574/wjaets.2024.11.1.0054}
}

@incollection{bharath2023,
  author = {Bharath and others},
  title = {Evaluating the Performance of Diverse Machine Learning Approaches in Stock Market Forecasting},
  booktitle = {Lecture Notes in Computer Science},
  year = {2023},
  pages = {255--264},
  doi = {10.1007/978-3-031-36402-0_23},
  url = {https://doi.org/10.1007/978-3-031-36402-0_23}
}

@inproceedings{rashid2010,
  author = {Rashid, A.},
  title = {Using a service oriented architecture for simulating algorithmic trading strategies},
  year = {2010},
  doi = {10.1145/1967486.1967650},
  url = {https://doi.org/10.1145/1967486.1967650}
}

@misc{quantconnect2024,
  author = {{QuantConnect}},
  title = {Overview - {QuantConnect.com}},
  year = {2024},
  url = {https://www.quantconnect.com/docs/v2/writing-algorithms/algorithm-framework/overview}
}

@incollection{sauberschwarz2017,
  author = {Sauberschwarz, L. and Weiss, L.},
  title = {How Corporations Can Win the Race Against Disruptive Startups},
  booktitle = {Springer Berlin Heidelberg},
  year = {2017},
  pages = {155--167},
  doi = {10.1007/978-3-662-49275-8_18},
  url = {https://doi.org/10.1007/978-3-662-49275-8_18}
}

@incollection{rakovic2018,
  author = {Rakovic, T. S.},
  title = {The Effect of Market Liquidity on the Company Value},
  booktitle = {Springer},
  year = {2018},
  pages = {183--196},
  doi = {10.1007/978-3-319-75907-4_12},
  url = {https://doi.org/10.1007/978-3-319-75907-4_12}
}

@article{cayirli2019,
  author = {Cayirli, O.},
  title = {The {Black-Litterman} Model: {Extensions} and Asset Allocation},
  journal = {SSRN},
  year = {2019},
  doi = {10.2139/ssrn.3464770},
  url = {https://doi.org/10.2139/ssrn.3464770}
}

@incollection{tzang2020,
  author = {Tzang, S.-W. and Hung, C.-H. and Tsai, Y.-S. and others},
  title = {{Black-Litterman} Model and Momentum Strategy: {Evidence} of Taiwan Top 50 {ETF}},
  booktitle = {Springer},
  year = {2020},
  pages = {490--497},
  doi = {10.1007/978-3-030-50399-4_47},
  url = {https://doi.org/10.1007/978-3-030-50399-4_47}
}

@incollection{alwateer2023,
  author = {Alwateer, M. M. and Atlam, E. and Farsi, M. and others},
  title = {Hidden {Markov} Models for Pattern Recognition},
  booktitle = {IntechOpen},
  year = {2023},
  doi = {10.5772/intechopen.1001364},
  url = {https://doi.org/10.5772/intechopen.1001364}
}

@article{qamar2023,
  author = {Qamar, R. and Zardari, B. A.},
  title = {Artificial Neural Networks: {An} Overview},
  journal = {Mesopotamian Journal of Computer Science},
  year = {2023},
  pages = {130--139},
  doi = {10.58496/mjcsc/2023/015},
  url = {https://doi.org/10.58496/mjcsc/2023/015}
}

@incollection{kwon2022,
  author = {Kwon, H. and others},
  title = {Distillation Column Temperature Prediction Based on Machine-Learning Model Using Wavelet Transform},
  booktitle = {Computer Aided Chemical Engineering},
  publisher = {Elsevier},
  year = {2022},
  pages = {1651--1656},
  doi = {10.1016/b978-0-323-85159-6.50275-x},
  url = {https://doi.org/10.1016/b978-0-323-85159-6.50275-x}
}

@article{zillich2009,
  author = {Zillich, R. and Chin, S. and Mayrhofer, J.},
  title = {Extrapolated High-Order Propagators for Path Integral {Monte Carlo} Simulations},
  year = {2009},
  doi = {10.48550/arxiv.0907.3495},
  url = {https://doi.org/10.48550/arxiv.0907.3495}
}

@article{wu2024,
  author = {Wu, L. and Picek, S. and Perin, G.},
  title = {I Choose You: {Automated} Hyperparameter Tuning for Deep Learning-Based Side-Channel Analysis},
  journal = {IEEE Transactions on Emerging Topics in Computing},
  year = {2024},
  volume = {12},
  pages = {546--557},
  doi = {10.1109/tetc.2022.3218372},
  url = {https://doi.org/10.1109/tetc.2022.3218372}
}

@article{ong2011,
  author = {Ong, H. F. and Ahmad, A. M.},
  title = {Malay Language Speech Recogniser with Hybrid Hidden {Markov} Model and Artificial Neural Network ({HMM/ANN})},
  journal = {International Journal of Information and Education Technology},
  year = {2011},
  pages = {114--119},
  doi = {10.7763/ijiet.2011.v1.19},
  url = {https://doi.org/10.7763/ijiet.2011.v1.19}
}

\end{document}